# Agriculture Credit and Economic Growth in Bangladesh: A Time Series Analysis

Md. Toaha, Laboni Mondal

**Abstract:** The paper examined the impact of agricultural credit on economic growth in Bangladesh. The annual data of agriculture credit were collected from annual reports of the Bangladesh Bank and other data were collected from the world development indicator (WDI) of the World Bank. By employing Johansen cointegration test and vector error correction model (VECM), the study revealed that there exists a long run relationship between the variables. The results of the study showed that agriculture credit had a positive impact on GDP growth in Bangladesh. The study also found that gross capital formation had a positive, while inflation had a negative association with economic growth in Bangladesh. Therefore, the government and policymakers should continue their effort to increase the volume of agriculture credit to achieve sustainable economic growth.

## I. Introduction

Bangladesh is a small country of south Asia with a population of 200 million, around 60.20% of them live in the rural area who don't not have access to the formal financial system. Over the last few decades, Bangladesh have achieved astonishing growth of GDP from 5.6% in 1990 to 7.2% in 2022. Researchers and experts have identified several factors behind this success. Export earnings (Balassa, 1978; Jiban et al., 2022), remittance inflows (Cazachevici et al., 2020; Jude et al., 2019), foreign direct Investments (Alvarado et al., 2017; Osei & Kim, 2020), foreign grant and aid (Das & Sethi, 2020; K. M. S. Islam & Biswas, 2023), are contributing for this remarkable growth. Many of them also believe Bangladesh couldn't have achieved its full potentials due to some problems, and barrier of access to formal financial system by poor people is one of them. To eliminate this problem, Bangladesh Bank (the central bank of Bangladesh) and the government of Bangladesh have taken so many steps. Agricultural credit is one of them, and it can be defined as a small loan for poor entrepreneurs from remote areas. Usually, these underprivileged group does not have enough collateral to get loans from banks and financial institutions. The central bank of Bangladesh, i.e. the Bangladesh Bank has formulated and implemented relevant policies and guidelines so that this section of people of the economy have access to formal credit market and get loans to start their startup. Due to these policy initiatives, these entrepreneurs can apply for loans from lenders and start their own businesses. Some financial institutions have developed several strategies, including providing small loans to rural poor people without guarantees. These loans can be repaid in predetermined installments. The borrower is organized into groups, thereby reducing the risk of default. These credits also help to disseminate valuable information about borrowers and their standard of living.

Adequate availability of fund is an important requirement for investors, especially in times of scarce resources and uncertainty. Convenient and safe savings facilities may be even more important to smooth consumption over time. The lack of savings facilities also forces families to rely on inefficient, uncomfortable, and expensive alternatives. For meeting the need of funds of these small entrepreneurs, Bangladesh Bank issued circulars and directed the banks and non-financial institutions operating in Bangladesh to disburse loans to agriculture sector without collateral or against minimum collateral value. This collateral-free financing offers a new opportunity for bank and financial institutions to invest in small risks. Previous experiences of financial institutions shows that small loans to the poor have proved to be appropriate for rural development. Thus, if they can effectively engage, agricultural credit will accelerate Bangladesh's rural development activities.

Currently, agriculture credit is not a new phenomenon in Bangladesh as it has started long ago. The amount of agriculture credit disbursed by banks and financial institutions has increased from 15.17 billion BDT in 1997 to 255.11 billion BDT in the year 2021, and this special type of credit has been helping to eradicate poverty and contributing to the economic growth especially in rural area of Bangladesh. Therefore, the objective of this study is to assess the effect of agriculture credit on economic growth in Bangladesh.

## II. Literature Review

There is dearth of empirical literatures on exploring the relationship between agriculture credit and economic growth in Bangladesh as well as around the world. Few studies have been conducted in this regard across the world and these studies mainly tested the hypothesis that whether there is a positive relationship between agriculture credit and economic growth. In the next part, I reviewed some literature related in this context.

M. Islam (2020) examined the impact of agriculture credit on agriculture productivity in Bangladesh using data from 2000 to 2019. Using autoregressive distributive lag (ARDL) model, the study revealed that there exists a short run and long run relationships between agriculture credit and agriculture productivity in Bangladesh. Finally, the study concluded that policymakers should take appropriate step to increase agriculture credit to enhance agriculture productivity.

Akmal et al. (2012) analyzed the impact of agriculture credit on economic growth in Pakistan during 1970-2010. By employing econometric technique, the study showed that agriculture credit had a significant positive impact on gross domestic product of agriculture sector in Pakistan. The study also revealed a positive correlation between agriculture labor force and agriculture GDP. Chakraborty & Shukla (2020) explored the effect of agriculture credit on economic development in India. Using a panel data set of state level from fiscal year 1995-96 to 2018-19, the study showed that a 10% percent increase in agriculture credit would increase agriculture GDP by 2.7%, which is statistically significant. The study also revealed that agriculture credit had a positive relationship with rural expenditure and rural infrastructure.

Hartarska et al. (2015) studied the association between agriculture credit and economic growth in USA. Using the state level and regional level data for the period 1991-2010, the study found that agriculture credit had a positive influence on rural GDP growth in USA. Furthermore, the findings also indicated that an increase in agricultural loans by one percentage results into 10% higher growth in agriculture GDP in state level, while that is 1.2 % in regional level.

Ayeomoni & Aladejana (2016) analyzed the association between agricultural credit and economic growth in Nigeria. By using the annual data from 1986 to 2014, the study showed that there was a short run as well as a long run relationship between agricultural credit and economic growth in Nigeria during the study period. The study also revealed that real exchange rate had a positive, while inflation had a negative effect on growth.

Narayanan (2016) explored the impact of agricultural credit on agricultural GDP in India. By using a panel data set of different states of India for the period between 1995-96 and 2011-12, the study found a weak relationship between growth of agricultural credit and growth of GDP in agriculture sector, whereas agricultural credit had a positive association with inputs.

## III. DATA AND METHODOLOGY

For conducting the study, I have derived annual data during 1997-2021 from various secondary sources. The data of agriculture credit has been collected from annual reports of different years of Bangladesh Bank (*Annual Report, Bangladesh Bank*, 2021), which is the variable of my interest in this study. I have used the gross domestic product (GDP) as the proxy for economic growth. Gross capital formation (GCF) is also another factor, which have profound effect on economic growth (Solow, 1962) and inflation is also another macroeconomic factor that determines the business environment of a country, which defines the growth trajectory of any economy (Barro, 1995; Biswas, 2023).

Based on the discussion above, I have estimated the following equation to test the hypothesis that agriculture credit has a positive relationship with economic growth in Bangladesh:

$$GDP_t = \beta + \alpha_1 \, AC_t + \alpha_2 \, GCF_{it} + INF_i + \varepsilon_t \qquad (1)$$

Where, GDP = gross domestic product
AC = Agriculture credit
GCF = Gross Capital formation
INF = Inflation and
$\varepsilon_t$ = error term

## IV. EMPIRICAL RESULTS

### A. Trend of Agriculture Credit and GDP

The below Fig. presents the graph of agriculture credit and GDP of Bangladesh.

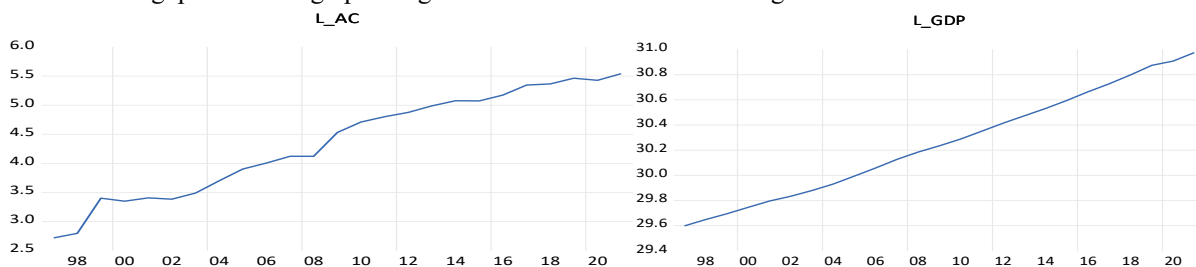

**Fig 1: Graph of Agriculture credit and GDP**

From the above fig. 1, we see that these variables have clear trend, which indicates that the variables might have unit root. Therefore, in the following part, I conducted ADF test to check for the stationarity of variables.

## B. ADF Test

Time series data run the risk of being non-stationary, and the results from non-stationary data might be biased and inconsistent. Hence, the first step of this sort of analysis is checking for the stationarity of the series. At this stage, I conducted the widely used Augmented-Dicky Fuller (ADF) test and the results are presented in Table I.

TABLE I: ADF TEST

| Variable | Parameter | ADF Statistics | P-Value | Decision |
|---|---|---|---|---|
| l_gdp | Level | -1.960213 | 0.5925 | I(1) |
|  | First diff. | -4.361670 | 0.0112 |  |
| l_ac | Level | -2.350178 | 0.3937 | I(1) |
|  | First diff. | -5.981250 | 0.0003 |  |
| L_gcf | Level | -3.090231 | 0.1328 | I(1) |
|  | First diff. | -4.787654 | 0.0036 |  |
| l_inf | Level | -2.387716 | 0.1554 | I(1) |
|  | First diff. | -4.388572 | 0.0024 |  |

From the above table I, it is evident that all variables are non-stationary at levels. However, after first differencing, all variables became stationary. Therefore, the variables are integrated of order one.

## C. Lag Length Selection

Before choosing Vector Auto-regression (VAR) or Vector Error Correction Model (VECM), we must find the optimal lag, which is shown below Table II:

TABLE II: OPTIMAL LAG SELECTION

| lags | AIC | SC | HQ |
|---|---|---|---|
| 1 | -4.446953 | -4.249476 | -4.397288 |
| 2 | -13.08429 | -12.09691* | -12.83597 |
| 3 | -13.32152* | -11.54423 | -12.87454* |

AIC: Akaike information criterion
SC: Schwarz information criterion
HQ: Hannan-Quinn information criterion

AIC and HQ criterion show that the optimal lag is 2, whereas the SC criterion shows that optimal lag is 1. Therefore, while estimating VAR/VECM in the next stage, I will choose optimal lag 2.

## D. Johansen Cointegration Test

We need to check for any probable long-term association or cointegration between the variables before choosing either VAR or vector error-correction model (VECM) model. Any probable cointegration would indicate that the series move together in the long run. In this section, I used the popular Johansen cointegration test, whose results is shown below Table III:

TABLE IV: JOHANSEN TEST

| Rank | Trace Test | | Max-Eigen Test | |
|---|---|---|---|---|
|  | Test statistics | p-value | Test statistics | p-value |
| None* | 82.42512 | 0.0000 | 0.888310 | 0.0000 |
| At most 1* | 0.691387 | 0.0146 | 0.691387 | 0.0100 |
| At most 2 | 0.238341 | 0.4302 | 0.238341 | 0.6145 |

Both Trace and Max-Eigen tests indicate that there are 2 cointegrating equations between the variables. It means that there exists a long-run relationship between the variables. The long-run equation is as follows:

$$l\_gdp = 0.111100 * l\_ac + 0.562799 * l\_gcf - 0.126099 * l\_inf \qquad (2)$$

Therefore, in the long run both agricultural credit and gross capital formation had a positive impact, whereas inflation had a negative impact on GDP.

*E. Vector Error Correction Model (VCEM) Estimation*

In the previous section, we found that that there exists long-run relationships between the variables. Therefore, I employed VECM model to reveal the short-run relationship among the variables. The result of VECM estimation is presented below Table IV:

TABLE IV: VECM ESTIMATION

| Variables | Coefficient | Std. Error | t-statistics | Significance |
|---|---|---|---|---|
| $\alpha_0$ | 0.053163 | 0.01519 | 3.49961 | *** |
| $d\_l\_gdp(1)$ | -0.087885 | 0.26625 | -3.33008 | *** |
| $d\_l\_ac(1)$ | 0.018130 | 0.01877 | 0.96602 | |
| $d\_l\_gcf(1)$ | 0.092533 | 0.15621 | 0.59237 | |
| $d\_l\_inf(1)$ | -0.003774 | 0.00754 | -0.50083 | |
| $EC_{t-1}$ | -0.233184 | 0.14104 | -2.65333 | *** |
| $EC_{t-2}$ | -0.034734 | 0.01524 | -2.89034 | *** |
| R-squared | | | 0.309407 | |
| Adjusted R-squared | | | 0.050434 | |
| F-statistic | | | 1.194748 | |

Note: *** means 5% significance levels

The result shows that the correction term $EC_{t-1}$ and $EC_{t-2}$ are negative, which are statistically significant at 5% significance level. The coefficients of error correction terms are -0.233184 and -0.034734, it means that previous year's disequilibrium corrected by 23.31% and 3.47 % respectively in the current period. The short-term coefficients of agriculture credit and gross capital formation are also positive, while the short term coefficient of inflation is negative. But these coefficients are not statistically significant.

## V. DIAGNOSTIC TESTS

*A. Heteroskedasticity Test*

The result of the Heteroskedasticity test is presented below table V:

TABLE V: HETEROSCEDASTICITY TEST

| Test Statistics | p-value |
|---|---|
| Chi-sq = 115.8519 | 0.1328 |

The null hypothesis of heteroskedasticity test is that the data is homoscedastic. We see that the p-value is 0.1328, it means that we fail to reject the null hypothesis. Therefore, the error terms are homoscedastic.

*B. Normality Test*

The result of normality test is given below.

TABLE VI: NORMALITY TEST

| Component | Skewness | Chi-sq | df | Prob.* |
|---|---|---|---|---|
| 1 | -0.767024 | 2.255247 | 1 | 0.1332 |
| 2 | 0.073798 | 0.020877 | 1 | 0.8851 |
| 3 | 0.546343 | 1.144213 | 1 | 0.2848 |
| 4 | -0.203307 | 0.158446 | 1 | 0.6906 |
| Joint | | 3.578783 | 4 | 0.4660 |
| Component | Kurtosis | Chi-sq | df | Prob. |
| 1 | 3.105327 | 0.010632 | 1 | 0.9179 |
| 2 | 2.266203 | 0.516023 | 1 | 0.4725 |
| 3 | 2.353823 | 0.400147 | 1 | 0.5270 |
| 4 | 2.192007 | 0.625650 | 1 | 0.4290 |
| Joint | | 1.552451 | 4 | 0.8173 |

| Component | Jarque-Bera | df | Prob. |
|---|---|---|---|
| 1 | 2.265879 | 2 | 0.3221 |
| 2 | 0.536900 | 2 | 0.7646 |
| 3 | 1.544360 | 2 | 0.4620 |
| 4 | 0.784096 | 2 | 0.6757 |
| Joint | 5.131234 | 8 | 0.7435 |

From the above table, we see that all p-value of all the tests i.e. skewness, kurtosis and Jarque-Bera are greater than 5%, it means that we fail to reject the null hypothesis that the errors are multivariate normal. Therefore, the errors are normally distributed.

## VI. Conclusion

Bangladesh is an agriculture-based country, whose around 40% of its total population are directly and indirectly engaged with agriculture sector. After independence, the agricultural sector was the main engine of Bangladesh economy. At that time, agriculture sector's contribution to GDP was around 60%. Agriculture in Bangladesh is vital for the livelihoods of people, employment and contribution to GDP; But its contribution to GDP has declined over the past decade, from 17% in 2010 to 11.2 % in 2022. Researchers and experts identified several factors that are responsible for the decline of agriculture sector's contribution to GDP. Change of occupation, inadequate funding facilitates, urbanization, climate change etc. are responsible for this sharp decline of contribution of agriculture sector to GDP. As a result, the government and the central bank of Bangladesh have taken so many initiatives to revive the agriculture sector. Directing banks and financial institutions to give agriculture credit is one of the significant initiatives that have been taken by the central bank, and as result the amount of agriculture credit has increased considerably during the last decade.

The purpose of this study is to examine the impact of agriculture credit on economic growth in Bangladesh. For conducting the study, I collected the data related to agriculture credit during 1997-2021 from the Annual report of Bangladesh Bank. Other data of this study were collected from world development index (WDI) of World Bank. By employing Johansen cointegration test, I found that there exists a long run equilibrium relationship between agriculture credit, GDP, gross capital formation and inflation in Bangladesh. The study also revealed that agriculture credit had a positive effect on economic growth in Bangladesh. Finally, using VECM technique the study found that previous year's disequilibrium adjusted at rate around 23.31% and 3.47 % respectively in the present year. The findings of the study can be used by policymakers and the government to formulate effective policies related to agriculture credit to achieve sustainable economic growth.